\documentclass[10pt]{JHEP3}

\usepackage{epsfig,multicol,bbm}
\usepackage{amsmath}
\usepackage{graphicx}
\usepackage{dcolumn}
\usepackage{bm}
\usepackage{amssymb}
\usepackage{latexsym}

\newcommand\fverb{\setbox\fverbbox=\hbox\bgroup\verb}
\newcommand\fverbdo{\egroup\medskip\noindent%
            \fbox{\unhbox\fverbbox}\ }
\newcommand\fverbit{\egroup\item[\fbox{\unhbox\fverbbox}]}
\newbox\fverbbox

\title{Is Non-minimal Inflation Eternal?}

\author{Chao-Jun Feng\\
    Shanghai United Center for Astrophysics(SUCA), Shanghai Normal University,\\
    100 Guilin Road, Shanghai 200234, P.R.China\\
    E-mail: \email{fengcj@shnu.edu.cn}}

\author{Xin-Zhou Li\\
    Shanghai United Center for Astrophysics(SUCA), Shanghai Normal University,\\
    100 Guilin Road, Shanghai 200234, P.R.China\\
    E-mail: \email{kychz@shnu.edu.cn}}

\received{\today}       
\accepted{\today}       

\abstract{The possibility that the non-minimal coupling inflation could be eternal is investigated. We calculate the
quantum fluctuation of the inflaton in a Hubble time and find that it has the same value as in the minimal case in the
slow-roll limit. Armed with this result, we have studied some concrete non-minimal inflationary models including the
chaotic inflation and the natural inflation while the inflaton is non-minimally coupled to the gravity and we find that
these non-minimal inflations could be eternal in some parameter regions. }


\begin{document}

\section{Introduction}
Inflation has been remarkably successful in explaining the properties of the universe and the origin of the primordial
perturbation \cite{inflation}, which is thought of as the seed of the large scale structures. In the context of new
inflation, the early inflating universe is  driven by a scalar field called inflaton, which is at first living at an
unstable state like the false vacuum state, then it slowly rolls down to a stable state like the true vacuum state. An
interesting phenomenon in the inflationary scenario is the eternal inflation, which means the inflation never ends.

The eternal new inflation was first discovered by Steinhardt \cite{steinhardt-nuffield}, and Vilenkin
\cite{Vilenkin:1983xq} showed that new inflationary models are generically eternal. As we known, the decay of the false
vacuum  to a true vacuum is an exponential process, however, it also exponentially expands when it decays during the
inflation time. In fact, the rate of exponential expansion is always much faster than the rate of exponential decay in
any successful inflationary models. Therefore, the false vacuum never disappears and the total volume of the false
vacuum grow exponentially with time when inflation starts, see ref \cite{Guth:2007ng}.

Actually, some inflationary models like chaotic inflation does not have a false vacuum state, but it can also be
eternal, which is called slow-roll eternal inflation \cite{Linde:1986fd}. In these models, the inflaton is classically
rolling down the hill, and the change in the field during some time interval is influenced by the quantum fluctuations,
which can drive the field upward or downward relative to the classical trajectory. There is always some probability
that the classical evolution is smaller than the quantum fluctuations, then the inflaton will fluctuate up and not
down. Therefore, this process will continues forever and inflation will never ends. The recent progress on eternal
inflation, see ref. \cite{recent pro eternal} and for recent reviews see ref. \cite{Guth:2007ng, Linde:2007fr}.

There is a class of inflationary models called non-minimal inflations, in which the inflaton ($\varphi$) is
non-minimally coupled to the gravity and terms like $f(R,\varphi)$ are introduced in the effective action, where $R$ is
the Ricci scalar. The cosmological effects of the non-minimal inflationary models have been well studied. It shows that
the power spectrum is generally blue, and the problem of getting a running spectral index is eased in the non-minimal
inflation \cite{miaoLi}. For other recent studies on the non-minimal inflation, see ref. \cite{recent nonminimal} .
However, the eternal property of them has not studied in these literatures. In this paper, we have investigated that
whether the non-minimal coupling inflation could be eternal or not, and we focus on the simplest model with  $ f =
\zeta R\varphi^2$ .

The paper is organized as follows. In Section \ref{s bd}, we briefly review the background dynamics for the inflaton
and the metric, impose the slow-roll condition and define three slow-roll parameters including the ordinary two and a
new one. In Section \ref{flu inf}, we calculate the quantum fluctuation and the classical motion  of the inflaton
during a Hubble time. Then, in Section \ref{example}, we focus on some concrete non-minimal inflation models and find
that they could be eternal in some parameter regions. In the last section, we give some discussion the conclusions.

\section{Background dynamics }\label{s bd}

The Lagrange density of the inflaton non-minimally coupled to gravity is given by
\begin{equation}
   \mathcal{L} = -  \frac{1}{2}g^{\mu\nu}\partial_\mu\varphi\partial_\nu\varphi -
    V(\varphi)  - \frac{1}{2}f(R)\varphi^n \,,
\end{equation}
where $f$ is a function of the scalar curvature. The total action reads
\begin{equation}\label{action}
    S = \frac{1}{2}\int d^4x \sqrt{-g}~R
    + \int d^4x \sqrt{-g}~ \mathcal{L} \,,
\end{equation}
where we have set $8\pi G = M_p^{-2} = 1$ . In a flat FRW universe with the unperturbed metric
\begin{equation}\label{metric}
    ds^2 = -dt^2 + a^2(t)\left(dx^2 + dy^2 + dz^2\right) \,,
\end{equation}
and the scalar curvature reads
\begin{equation}\label{scalar cur}
    R = 6\left(\dot H + 2H^2\right)\,,
\end{equation}
where $H = \dot a/a$ is the Hubble parameter and the over dot denotes the derivative with respect to the co-moving time
$t$. By varying the action (\ref{action}) we obtain the field equations
\begin{eqnarray}
  3H^2 &=& \frac{1}{2}\dot\varphi^2 + V + \frac{1}{2}f\varphi^2 + 3H^2\left[ \frac{d}{dt}\left(\frac{f'\varphi^2}{H}\right) -
    f'\varphi^2\right] \,, \label{Fried1}\\
  -2\dot H &=& \dot\varphi^2 + H^3 \frac{d}{dt}\left(\frac{f'\varphi^2}{H^2}\right)- \frac{d^2}{dt^2}\bigg(f'\varphi^2\bigg) \,,
\end{eqnarray}
where prime denotes the derivative with its argument, i.e. $f' = df/dR$ and $V' = dV/d\varphi$ in the following E.O.M.
of the scalar field
\begin{equation}\label{eomp1}
    \ddot \varphi + 3H \dot\varphi + V' + f\varphi = 0 \,.
\end{equation}
To get an enough long time inflation, we impose the following slow-roll conditions
\begin{equation}
   |\dot H| \ll H^2 \,, \quad  |\ddot \varphi| \ll 3H|\dot\varphi| \,,
\end{equation}
which leads to
\begin{equation}\label{cond1}
 |f'\varphi^2 \dot H| \ll V
\,,\quad \bigg|V'' + f - \frac{3Hf'\dot R}{f}\frac{f\varphi}{V'+f\varphi}\bigg| \ll 9H^2 \,,
\end{equation}
and we also assume $\dot\varphi^2/2\ll V$ to simplify the following calculation. Moreover, if $f$ is a monomial of $R$,
e.g. $f\sim R^n$, then $3H (\log f)'\dot R \approx 6n\dot H $. Therefore, in the case of $|f\varphi|\ll |V'|$ or
$|f\varphi|\gg |V'|$, the third term on the L.H.S. of the second equation in (\ref{cond1}) can be neglected and it
simplifies to
\begin{equation}\label{cond2}
|f'\varphi^2 \dot H| \ll V \,, \quad \bigg|V'' + f \bigg| \ll 9H^2\,.
\end{equation}
With these conditions in mind, the equations of motion are simplified
\begin{eqnarray}
  3H^2 &=& \frac{1}{2}\left(f-6f'H^2\right)\varphi^2 + V \,, \label{Fried1}\\
  3H \dot\varphi &=& - V' \,,  \quad |f\varphi|\ll |V'|\quad(\text{Case I.}~ ) \label{eomp21}\,,\\
  3H \dot\varphi &=& - f\varphi \,, \quad |f\varphi|\gg |V'|\quad(\text{Case II.})\label{eomp22}\,.
\end{eqnarray}
The traditional the dimensionless slow-roll parameters are given by
\begin{equation}\label{slow para}
   \epsilon = -\frac{\dot H}{H^2} \,, \quad \eta = M_p^2\frac{V''}{V} \,,
\end{equation}
and we define another new a new dimensionless slow-roll parameter as
\begin{equation}\label{slow para2}
    \Delta \equiv M_p^2 \frac{f}{V} \,,
\end{equation}
where we have recovered the Planck mass to indicate these parameters being dimensionless. Then the slow-roll conditions
(\ref{cond2}) becomes
\begin{equation}\label{cond3}
    \epsilon \Delta \varphi^2\ll1 \,, \quad \eta + \Delta \ll1
\end{equation}
where we have used $f'\sim f/R$ and $V'\sim V/\varphi$. Therefore, if $\epsilon , \eta ,\Delta\ll1$, the slow-roll
conditions in (\ref{cond2}) is met . In the following, we will consider a simple class of model with $f = \xi R$. Thus,
the Friedmann equation (\ref{Fried1}) becomes
\begin{equation}\label{Fried mod1}
    3H^2 = \frac{V}{1-\xi\varphi^2} \,,
\end{equation}
and the new slow-roll parameter is
\begin{equation}\label{del1}
    \Delta = \frac{4\xi(2-\epsilon)}{(1-\xi\varphi^2)}  \,.
\end{equation}
Therefore, $\Delta\ll1$ requires $\xi\ll1$ in this model. In the case of $|f\varphi|\ll |V'|$, i.e.$ \Delta \varphi^2
\ll1$, the slow-roll parameter is
\begin{equation}
    \epsilon  = \frac{V'^2}{2V^2} \bigg[1-\left(1-\frac{2V}{V'\varphi}\right)\xi\varphi^2\bigg] \,,
\end{equation}
while in the case of $|f\varphi|\gg |V'|$, i.e.$ \Delta \varphi^2 \gg1$, it becomes
\begin{equation}
    \epsilon  = \frac{f\varphi V'}{2V^2} \bigg[1-\left(1-\frac{2V}{V'\varphi}\right)\xi\varphi^2\bigg] \,.
\end{equation}
It should be noticed that in the latter case, the condition $ \Delta \varphi^2 \gg1$ means we must have large field
inflation, i.e. $\varphi^2\gg M_p^2$, and in both case, $\xi\varphi^2 \sim \mathcal{O}(1)$ is required, otherwise it
reduces to the minimal coupling case. Therefore, in the following, we will only consider large field inflation.

\section{Fluctuation and motion of the inflaton}\label{flu inf}
In order to calculate the quantum fluctuation of the inflaton, we should expand the action (\ref{action}) to the second
order. The action approach guarantees the correct normalization for the quantization of fluctuations and it is
convenient to work in the ADM formalism. We write the metric as
\begin{equation}
    ds^2 = -N^2dt^2 + h_{ij}(dx^i + N^idt)(dx^j + N^jdt) \,,
\end{equation}
where $N$ is the lapse function and $N^i$ is the shift vector and the action (\ref{action}) becomes
\begin{equation}\label{action2}
    S = \frac{1}{2} \int dt dx^3 \sqrt{h}\bigg[NR^{(3)} + N^{-1}\left(E_{ij}E^{ij}-E^2\right)
    +N^{-1}\left(\dot\varphi - N^i\partial_i\varphi\right)^2 -
    Nh^{ij}\partial_i\varphi\partial_j\varphi-N\left(2V+f\varphi^2\right)\bigg]\,,
\end{equation}
where $h=\det h_{ij}$ and the symmetric tensor $E_{ij}$ is defined as
\begin{equation}
    E_{ij} = \frac{1}{2}\bigg(\dot h_{ij}-\nabla_iN_j-\nabla_jN_i\bigg) \,, \quad E = E^i_{~i} \,.
\end{equation}
Here $R^{(3)}$ is the three-dimensional Ricci curvature which is computed from the metric $h_{ij}$. Note that the
extrinsic curvature is $K_{ij}=E_{ij}/N$. We will work in the spatially flat gauge and neglect the tensor
perturbations:
\begin{equation}\label{flat gauge}
    \varphi(t,x) = \bar\varphi(t) + \delta\varphi(t,x) \,, \quad h_{ij}=a^2\delta_{ij} \,,
\end{equation}
where $\bar\varphi(t)$ is the background value of the scalar field and $\delta\varphi$ is a small fluctuation from the
background value. In the ADM formalism, one can think of $N$ and $N^i$ as Lagrange multipliers, and to get the action
for $\zeta$, one need to solve the constraint equations for $N$ and $N^i$ and plug the result back in the action.

The equations of motion for $N^i$ and $N$ are the momentum and hamiltonian constraints
\begin{eqnarray}
  \nabla_i\bigg[(1-f'\varphi^2)N^{-1}\left(E^i_j - \delta^i_jE\right)\bigg] -N^{-1}\left(\dot\varphi-N^i\partial_i\varphi\right)\partial_j\varphi  &=& 0 \,,\label{f constr1}\\
  R^{(3)} -(1-2f'\varphi^2) N^{-2}\left(E_{ij}E^{ij}-E^2\right) -N^{-2}\left(\dot\varphi-N^i\partial_i\varphi\right)^2- 2V-f\varphi^2 -h^{ij}\partial_i\varphi\partial_j\varphi&=& 0
  \,,\label{f constr2}
\end{eqnarray}
By decomposing $N^i$ into $N^i = \partial^i\psi + N^i_T$ where $\partial_iT^i_T=0$ and $N=1+N_1$, where
$N_1,N^{i}_T,\psi\sim \mathcal{O}(\delta\varphi)$, and plugging these expansion into the eqs. (\ref{f constr1}) and
(\ref{f constr2}) for $N$ and $N^i$, one can obtain the solutions in the first order of $\zeta$. In the case of $f =
\xi R$, we get the first order solutions in Appendix \ref{app mod1} , and in the following we will still use $\varphi$
to denote the background value $\bar\varphi$:
\begin{equation}\label{sol 11}
    N_1 = \frac{\delta\varphi}{1-\xi\varphi^2} \left(\frac{\dot{\varphi}}{2H} - 2\xi\varphi \right)\,, \quad N^{i}_T =0 \,,
\end{equation}
and
\begin{equation}\label{sol 12}
   (1-\xi\varphi^2) \partial^2\psi = N_1\frac{\dot{\varphi}^2}{2H}- \frac{\dot{\varphi}}{2H}\delta\dot\varphi
    -\left(\frac{3}{2} \frac{\dot{\varphi}}{H}- 3\xi\varphi + \frac{V'}{2H^2}\right)H\delta\varphi
\end{equation}
with suitable boundary conditions. We also get the exact background dynamic equation
\begin{equation}\label{bg1}
    3H^2 (1-\xi\varphi^2)= \frac{1}{2}\dot{\varphi}^2 + V \,,
\end{equation}
which is consistent with eq.(\ref{Fried mod1}) in the slow-roll limit. To find the quadratic action for
$\delta\varphi$, we need plug eqs.(\ref{sol 11}) and (\ref{sol 12}) in the action and expand it to the second order.
However, we can see that these expressions for $N$ and $N^i$ is are subleading  in slow-roll ($\dot{\varphi}^2\ll H^2$)
and large field ($\varphi^2\gg M_P^2$) inflation compared to $\delta\varphi$. So, it is enough to consider just the
action (\ref{action}) for $\delta\phi$ in the de Sitter background and we get the second order action
\begin{equation}
    S_2 = \frac{1}{2}\int d^4x a^3\bigg[ \delta\dot{\varphi}^2 - (\nabla{\delta\varphi})^2 - V''\delta\varphi^2 -
    12\xi H^2\delta\varphi^2\bigg] \,,
\end{equation}
and the perturbation equation is
\begin{equation}
    \delta\ddot\varphi_k + 3H\delta\dot\varphi_k+\frac{k^2}{a^2}\delta\varphi_k  =0 \,,
\end{equation}
where we have used $\eta \ll 1$ and $\Delta\ll1$ and $\delta\varphi_k$ is the Fourier transform of $\delta\varphi$.
Thus, the quantum fluctuation of $\delta_z$ in a Hubble time has the same value as in minimal case \cite{Guth:2007ng}
\begin{equation}
    \delta_q\varphi \approx \frac{H}{2\pi} \,,
\end{equation}
while usually the classical motion of the inflaton during one Hubble time is given by
\begin{equation}
    |\delta_c\varphi| \approx \dot{\varphi}H^{-1} \sim \frac{V'}{3H^2}\bigg(1+\Delta\varphi^2\bigg)\,.
\end{equation}
and the condition for eternal inflation to happen is roughly $\delta_q\varphi > |\delta_c\varphi|$.

\section{Eternal non-minimal inflation}\label{example}
In this section, we focus on some concrete large field inflation models. The first example is
\begin{equation}\label{potential}
    V(\varphi) = \frac{1}{2}m^2\varphi^2 \,,
\end{equation}
where $m$ is the mass of inflaton. Then, the condition of the eternal inflation is
\begin{equation}
     \frac{2(1+\Delta\varphi^2)}{\varphi}<\frac{H}{2\pi}\sim
     m\varphi \,.
\end{equation}
where we have used $\xi\varphi^2 \sim \mathcal{O}(1)$. Then, for the case of $\Delta\varphi^2\ll1$, it requires
\begin{equation}
    \varphi > \varphi_c = M_p\sqrt{\frac{M_p}{m}} \,, \quad \varphi\ll\Delta^{-1/2}M_p\,,
\end{equation}
where we have recovered the Planck mass and  $\varphi_c \gg M_p$ is the critical value of inflaton to be eternal. While
for the case of $\Delta\varphi^2\gg1$, it requires
\begin{equation}
\Delta < \frac{m}{M_p} \,, \quad \varphi\gg\Delta^{-1/2}M_p\,,
\end{equation}
to become eternal inflation. Therefore, in both case, the inflation could never be eternal if $m< \Delta M_p$.

The second example considered here is
\begin{equation}\label{potential2}
    V(\varphi) = \lambda \varphi^n \,,
\end{equation}
with $n>2$. Then, the condition of the eternal inflation is
\begin{equation}
     \frac{n(1+\Delta\varphi^2)}{\varphi}<\frac{H}{2\pi}\sim \lambda^{1/2}\varphi^{n/2} \,.
\end{equation}
Then, for the case of $\Delta\varphi^2\ll1$, it requires
\begin{equation}
    \varphi > \varphi_{c1} = n^{2/(n+2)}\frac{M_p}{\lambda^{1/(n+2)}} \,,  \quad \varphi\ll\Delta^{-1/2}M_p\,.
\end{equation}
So, if $\lambda < \Delta^{(n+2)/2}$, the inflation could never be eternal. While for the case of $\Delta\varphi^2\gg1$,
it requires
\begin{equation}
    \varphi > \varphi_{c2} = (n)^{2/(n-2)}\bigg(\frac{\Delta^2}{\lambda}\bigg)^{1/(n-2)} M_p\,, \quad \varphi\gg\Delta^{-1/2}M_p\,,
\end{equation}
to become eternal inflation and it is always an eternal inflation if $\lambda > \Delta^{(n+2)/2}$.

The above two examples  both belong to the class of the chaotic inflation, which is a typical large field inflation,
while the inflaton is non-minimal coupled to the gravity. Another elegant inflationary model is called the natural
inflation where the potential takes the following form
\begin{equation}\label{potential 3}
    V(\varphi) = V_0\bigg[\cos\left(\frac{\varphi}{g}\right)+1\bigg] \,,
\end{equation}
and this model can be of the small field or large field type depending on the parameter $g$. If $2\pi g >M_p$, namely
the periodicity of the inflaton larger than the Planck scale, it would be a large field inflation. Then, the condition
of the inflation being eternal is
\begin{equation}\label{ni1}
   F(\varphi)^{2/3} <  V(\varphi)\,, \quad \varphi\ll\Delta^{-1/2}M_p\,,
\end{equation}
for the case of $\Delta\varphi^2\ll1$, and
\begin{equation}\label{ni2}
    F(\varphi)^{2/3}\bigg(\Delta\varphi^2\bigg)^{2/3}< V(\varphi)\,, \quad \varphi\gg\Delta^{-1/2}M_p\,,
\end{equation}
for the case of  $\Delta\varphi^2\gg1$. Here we have defined the function
\begin{equation}
    F(\varphi)=-\frac{V_0}{ g} \sin\left(\frac{\varphi}{g}\right)\,.
\end{equation}

\begin{figure}[h]
\begin{center}
\includegraphics[width=0.6\textwidth]{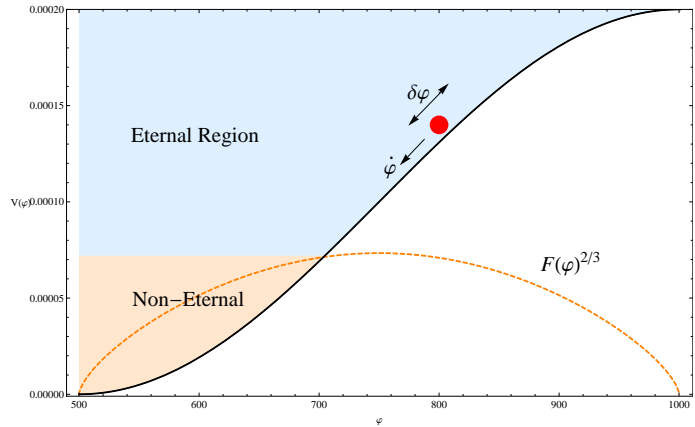}
\caption{\label{fig::case1} \textit{Case:}$\Delta\varphi^2\ll1$. The eternal and non-eternal regions are illustrated
with $V_0 = 10^{-4}$, and $2\pi g=10^3$ in the Planck units. }
\end{center}
\end{figure}

\begin{figure}[h]
\begin{center}
\includegraphics[width=0.6\textwidth]{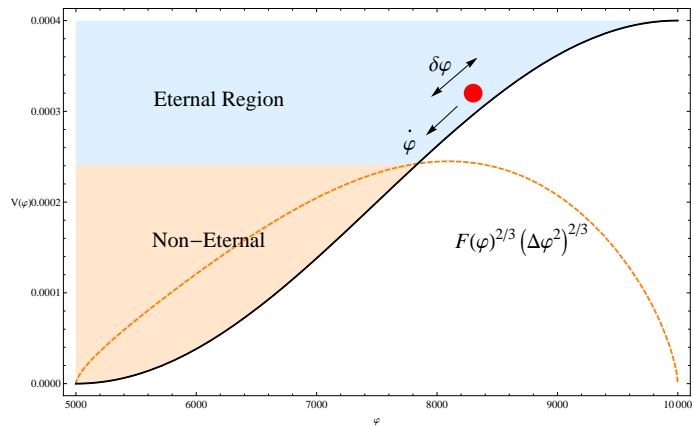}
\caption{\label{fig::case2}\textit{Case:}$\Delta\varphi^2\gg1$. The eternal and non-eternal regions are illustrated
with $V_0 = 2\times10^{-4}\,, 2\pi g=10^4$, and $\Delta = 10^{-6}$  in the Planck units.  }
\end{center}
\end{figure}

To illustrate the region in which the inflation could be eternal, one should solve eqs.(\ref{ni1}) and (\ref{ni2}).
Since they do not have analytical solutions, we have solved them numerically and illustrated the results in
Figs.(\ref{fig::case1}) and (\ref{fig::case2}) with particular parameters to give an example. It should be noticed
that, in Fig.(\ref{fig::case1}), $\varphi\ll\Delta^{-1/2}$, and if $\Delta$ is not small enough, there would be no
eternal region. For example, let $\Delta = 10^{-6}$, not all the areas in Fig.(\ref{fig::case1}) are valid since
$\varphi \ll 10^3$ is required. In other words, in the case of $\Delta\varphi^2\ll1$, the inflation could be eternal
only if $\Delta$ is much small. In Fig.(\ref{fig::case2}), we have set $\Delta = 10^{-6}$ for numerical calculation,
and then $\varphi \gg10^3$ is required in this case.

\section{Conclusions}

Many inflationary models have eternal properties like the new inflation and the chaotic inflation. In this paper, we
have studied the  eternal properties of  non-minimal inflationary models, in which there is a non-minimal coupling
between the inflaton and the gravity. We have calculated the quantum fluctuation of the inflaton at the first order,
and find that it has the same value as  in the minimal case, namely, the quantum fluctuation of inflaton in a Hubble
time is roughly proportional to the Hubble parameter during the inflation. If the quantum fluctuation overcome the
classical motion of the inflaton, the inflation could  never end.

We have studied some concrete non-minimal inflationary models including the chaotic inflation with power law potentials
and the natural inflation, derived the conditions under which the inflation could be eternal and found that in some
parameter spaces, it could be eternal inflation. Our results could be simply generalized to the case of $f \sim
R\varphi^n$, where $n=2$ is studied in this paper.

\acknowledgments This work is supported by National Science Foundation of China grant No. 10847153 and No. 10671128.

\appendix
\section{First order solution} \label{app mod1}
In this model, the constraint eqs. (\ref{f constr1}) and (\ref{f constr2}) becomes
\begin{eqnarray}
  \partial_i\bigg[(1-\xi\varphi^2)N^{-1}\left(E^i_j - \delta^i_jE\right)\bigg] -N^{-1}\left(\dot\varphi-N^j\partial_j\varphi\right)\partial_i\varphi  &=& 0 \,,\label{f1 constr1}\\
  (1-\xi\varphi^2)\bigg[R^{(3)} - N^{-2}\left(E_{ij}E^{ij}-E^2\right)\bigg] -N^{-2}\left(\dot\varphi-N^i\partial_i\varphi\right)^2- 2V -h^{ij}\partial_i\varphi\partial_j\varphi&=& 0
  \,,\label{f1 constr2}
\end{eqnarray}
with the ansatz
\begin{equation}\label{ansatz}
    N = 1+N_1  \,, \quad N^i = \partial^i\psi + N^i_{T} \,,\quad \partial_i N^i_{T} =0 \,,\quad N_i = h_{ij}N^j\,,
\end{equation}
and the spatially flat metric
\begin{equation}
    h_{ij} = a^2\delta_{ij} \,, \quad  h^{ij} = a^{-2}\delta^{ij} \,, \quad \sqrt{ h}=\sqrt{\det h_{ij}}=
    a^3 \,,\quad \Gamma^i_{ij}=0 \,, \quad R^{(3)}=0\,.
\end{equation}
After some length calculations, we find
\begin{equation}
    E^{ij}E_{ij}-E^2 =-6H^2+4H\partial^2\psi \,,
\end{equation}
and
\begin{equation}
    E^{i}_j-\delta^i_jE  = -2\delta^i_j H - \partial^i\partial_j\psi +\delta^i_j\partial^2\psi -\delta_{jl}\partial^iN^l_{T}
\end{equation}
Thus, eqs.(\ref{f constr1}) and (\ref{f constr2}) becomes
\begin{eqnarray}
  2H(1-\xi\bar\varphi^2)\partial_jN_1 + 4H\xi\bar\varphi\partial_j\delta\varphi
  -(1-\xi\bar\varphi^2)\delta_{jl}\partial^2N^l_{T}&=&\dot{\bar\varphi}\partial_i\delta\varphi \,, \label{A sol1}\\
  -\bigg(1-\xi\bar\varphi^2\bigg)4H\partial^2\psi-12H^2 \left(\frac{\dot{\bar\varphi}}{2H}\delta\varphi - \xi\bar\varphi\delta\varphi
  \right) &=& -2N_1\dot{\bar\varphi}^2+ 2\dot{\bar\varphi}\delta\dot\varphi+2V'\delta\varphi \,. \label{A sol2}
\end{eqnarray}
From eq.(\ref{A sol1}), one finds
\begin{equation}
    N_1 = \frac{\delta\varphi}{1-\xi\bar\varphi^2} \left(\frac{\dot{\bar\varphi}}{2H} - 2\xi\bar\varphi \right)\,, \quad \partial^2N^{i}_T =0 \,,
\end{equation}
so with appropriate choice of boundary conditions one can justifiably set $N^i_T = 0$. Plugging the solution of $N_1$
into eq.(\ref{A sol2}) we obtain the result (\ref{sol 12}).

\end{document}